\documentclass[showpacs,preprintnumbers,amsmath,amssymb,nofootinbib]{revtex4}
\usepackage{graphicx}
\usepackage{color, soul}
\usepackage{amsmath}
\usepackage{amsfonts}
\input epsf
\usepackage[pdftex,
            pdfauthor={A.A. Yurova, V. A. Yurov, A. V. Yurov},
            pdftitle={The Cauchy problem for the generalized hyperbolic Novikov-Veselov equation via the Moutard symmetries},
            pdfsubject={},
            pdfkeywords={Novikov-Veselov equation, Lax pair, Moutard symmetry, Cauchy problem},
            pdfproducer={Latex with hyperref},
            pdfcreator={pdflatex}]{hyperref}
\hypersetup{colorlinks=true, citecolor=blue, linkcolor=red}
\usepackage{enumerate}
\usepackage{scrextend}
\addtokomafont{labelinglabel}{\sffamily}

\newcommand{\beq}{\begin{equation}}
\newcommand{\beqn}{\begin{equation*}}
\newcommand{\enq}{\end{equation}}
\newcommand{\enqn}{\end{equation*}}
\newcommand{\eb}{{\rm e}}

\newcommand{\R}{{\mathbb R}}
\newcommand{\N}{{\mathbb N}}

\newcommand{\Ai}{\text{\rm Ai}}

\begin{document}
\title{The Cauchy problem for the generalized hyperbolic Novikov-Veselov equation via the Moutard symmetries}
\author{A. A. Yurova}%
\email{AIUrova@kantiana.ru}
\author{A.V. Yurov}%
\email{AIUrov@kantiana.ru}
\author{V.A. Yurov}%
\email{vayt37@gmail.com}
\affiliation{I. Kant Baltic Federal University, Department of Physics, Mathematics and IT, Al. Nevsky str. 14, Kaliningrad
236041, Russia}

\begin{abstract}
We begin by introducing a new procedure for construction of the exact solutions to Cauchy problem of the real-valued (hyperbolic) Novikov-Veselov equation which is based on the Moutard symmetry. The procedure shown therein utilizes the well-known Airy function $\Ai(\xi)$ which in turn serves as a solution to the ordinary differential equation $\frac{d^2 z}{d \xi^2} = \xi z$. In the second part of the article we show that the aforementioned procedure can also work for the $n$-th order generalizations of the Novikov-Veselov equation, provided that one replaces the Airy function with the appropriate solution of the ordinary differential equation $\frac{d^{n-1} z}{d \xi^{n-1}} = \xi z$.
\end{abstract}

\maketitle

\section{Introduction} \label{sec:Intro}
\allowdisplaybreaks

The Moutard transformation \cite{Moutard}  is a very interesting form of discrete symmetry of second-order linear equations with variable coefficients.
Consider the differential equation
\begin{equation}
\Delta\psi-U(\xi_1,\xi_2)\psi= \Delta\phi - U(\xi_1,\xi_2)\phi=0,
\label{25}
\end{equation}
where $\Delta$ is  the two-dimensional Euclidean Laplacian. Let $\psi=\psi(\xi_1,\xi_2)$ and $\phi=\phi(\xi_1,\xi_2)$ be two partial solutions of this equation, i.e. its solutions for two different Cauchy problems (with different initial conditions). The function $\phi$ we will call the {\it prop solution} because it will play the role of a foundation on which we will build the Moutard's mathematical apparatus. This apparatus, known as {\em the Moutard symmetry} is to be defined as the following transformation:
\begin{equation}
\psi\to\psi[1]=\frac{\theta[\psi,\phi]}{\phi},\qquad U\to
U[1]=U-2\Delta\ln\phi,
\label{26}
\end{equation}
where
\begin{equation}
\theta[\psi,\phi]=\int_{\Gamma}dx_{\mu}\varepsilon_{\mu\nu}
\left(\phi\partial_{\nu}\psi-\psi\partial_{\nu}\phi\right),
\label{27}
\end{equation}
and we used the following (standard) tensor notations: $\mu\in\{1,2\}$; $\partial_{\mu}=\partial/\partial \xi_{\mu}$; $\varepsilon_{\mu\nu}$ is a fully antisymmetric tensor with $\varepsilon_{12}=1$; and, as usual, we have a summation over the repeated indices. Note that the one-form being integrated in (\ref{27}) is a closed one when $\psi$ and $\phi$ are both solutions of (\ref{26}). Hence, the shape of the contour of integration $\Gamma$ in (\ref{27}) is irrelevant.

If we switch from $(\xi_1, \xi_2)$ to the cone variables $x$ and $y$, then the equation (\ref{25}) will take a new form:
\begin{equation}
\left(\frac{\partial^2}{\partial x\partial
y}+U(x,y)\right)\Psi(x,y)=0, \label{Wheeler-DeWitt-1}
\end{equation}

To define the Moutard symmetry in the new cone variables one has to override the closed one-form $\theta$. Namely, let $\Psi=\Psi(x,y)$ and $\Phi=\Phi(x,y)$ be two solutions of the (\ref{Wheeler-DeWitt-1}):
\begin{equation}
-\frac{1}{\Psi}\frac{\partial^2\Psi}{\partial x\partial
y}=-\frac{1}{\Phi}\frac{\partial^2\Phi}{\partial x\partial
y}=U(x,y). \label{solutions}
\end{equation}
Define a differential form $d\theta[\Psi;\Phi]$ such that
\begin{equation}
d\theta[\Psi;\Phi]= dx\left(\frac{\partial\Psi}{\partial x
}\Phi-\frac{\partial\Phi}{\partial x
}\Psi\right)-dy\left(\frac{\partial\Psi}{\partial y
}\Phi-\frac{\partial\Phi}{\partial y }\Psi\right), \label{form}
\end{equation}
and
\begin{equation}
\theta[\Psi;\Phi]=\int_{\Gamma}d\theta[\Psi;\Phi]. \label{theta}
\end{equation}
Note that since by definition both $\Psi$ and $\Phi$ are solutions of (\ref{solutions}), the one-form is closed too, i.e.
$$
\frac{\partial^2\theta[\Psi;\Phi]}{\partial x\partial
y}=\frac{\partial^2\theta[\Psi;\Phi]}{\partial y\partial x},
$$
and thus the shape of the contour of integration $\Gamma$ is irrelevant.

The Moutard symmetry  has the form
\begin{equation}
\Psi\to\Psi[1]=\frac{\theta[\Psi;\Phi]}{\Phi}, \label{Psi1}
\end{equation}
\begin{equation}
U\to U[1]=U+2\frac{\partial^2}{\partial x\partial y}\ln\Phi.
 \label{U1}
\end{equation}
It means that
\begin{equation}
-\frac{1}{\Psi[1]}\frac{\partial^2\Psi[1]}{\partial x\partial
y}=U[1](x,y). \label{dressed}
\end{equation}
It is worth pointing out at this step that
\begin{equation}
\Phi[1]=\frac{1}{\Phi}, \label{Phi1}
\end{equation}
rather then zero.

The Moutard symmetry in usual variables can be formally reduced in a one-dimensional limit to the Darboux transformation (originally introduced in 1882 in article \cite{Darboux1882} and further developed by Crum in \cite{Crum}). In other words, the Moutard symmetry group includes the Darboux transformation group as a subgroup. This is potentially a very significant observation since the group structure of the Darboux transformation  is associated with a subalgebra of the Kats--Moody algebra of the group SU(2) by means of the following commutation relation:
$$
\left[L_n^{\mu},L_m^{\nu}\right]=c_{\rho}^{\mu\nu}L^{\rho}_{m+n},
$$
where $c_{\rho}^{\mu\nu}$ are the structure constants of the group SU(2) (i.e. the group of unitary $2\times 2$ matrices with determinant 1, whose generators are the famous Pauli matrices.) \cite{Berezovoi87} . It is important to note, however, that while the Darboux transformation is indeed a limit case of the Moutard transformation in the one-dimensional limit, there still exist a crucial difference between the two. While both require two eigenfunctions to jump start the process of transformation, the Darboux transformation allows them to belong to {\em different eigenvalues}, as long as the prop function is positively defined everywhere (to ensure the regularity of a new dressed potential). In contrast to that, the Moutard symmetries demands a pair of functions to belong to the {\em same eigenvalue}, which in this paper for simplicity is set equal to zero.

Another interesting property of the Moutard symmetries is that they ((\ref{Psi1}), (\ref{Phi1}) , (\ref{U1}) or (\ref{26}), (\ref{27})) can be iterated several times, and the result of their application can be expressed via the corresponding Pfaffian forms \cite{Matveev91}. The Moutard symmetries have a number of interesting physical applications, ranging from the Maxwell equations in a 2D inhomogeneous medium \cite{Yurova} to the problems of quantum cosmology and the Wheeler-DeWitt equation \cite{Astashenok}, as well as the quantum (phantom) modified gravity models \cite{Briscese07}. But the Moutard transformations are most effective in the theory of integrable evolutionary equations in (1 + 2) dimensions.

In 1984 two Soviet mathematicians, Sergey Novikov and Aleksander Veselov, made a very important discovery \cite{Novikov84}. At that time they were studying the class of two-dimensional Schr\"odinger operators $L$ with a property of a ``finite-zoneness w.r.t. a single energy level'', originally introduced in 1976 by Dubrovin, Krichever and Novikov in \cite{Dubrovin76}, and is a property of a Schr\"odinger operator $L$ whose Bloch functions (i.e. the eigenfunctions that $L$ shares with the periodic operators of spatial translations) on a single energy level is meromorphic (i.e. holomorphic everywhere except for a few isolated poles) on a Riemann surface of a finite genus (\cite{Dubrovin76}, see also \cite{Matveev76}). More specifically, Veselov and Novikov were interested in those ``finite-zone'' 2D Schr\"odinger operators $L$ that are simultaneously real and purely potential. For that end, a theorem has been proven that for eigenfunction $\psi$ of such an operator to be meromorphic everywhere except at two points and possess the required asymptotes (see \cite{Novikov84} for details), those eigenfunctions necessarily have to satisfy the following system of equations:
\beq \label{hierarchy}
\begin{split}
L  \psi &= \partial \bar\partial \psi + V \psi = 0, \\
\frac{\partial\psi}{\partial t_n} &= \left(A_n - \bar A_n\right)\psi,
\end{split}
\enq
where the overhead bar denotes complex conjugate, the operators
\beq \label{A_n}
\begin{split}
A_n &= \partial^{2n+1} + a_{2n-1} \partial^{2n-1} + ... + a_0, \\
\partial &= \frac{1}{2}\left(\frac{\partial}{\partial x} - i \frac{\partial}{\partial y}\right),
\end{split}
\enq
and the coefficients $a_j$  and $V$ are uniquely defined by the aforementioned asymptotes.

The pair of operators $L$ and $A$ coupled by the system \eqref{hierarchy} deeply resembles the famous Lax pair $\mathcal L, \mathcal A$ \cite{Lax76}: two time-dependent operators that satisfy the condition
\beqn
\frac{d\mathcal L}{dt}=[\mathcal{A, L}],
\enqn
which in turn produces a new $(1+1)$-dimensional differential equation -- with such notable examples as, Korteweg-de Vries (KdV), sine-Gordon and the nonlinear Schr\"odinger equations (for more details cf. \cite{Lax76} and \cite{Matveev91}, see also \cite{Yurov18}). However, in the case of \eqref{hierarchy} the proper ``entanglement'' between the $L$ and $A$ was a tad more complex, and required an additional operator $B_n$ (similar in form to $A_{n-1}$ albeit with its own coefficients):
\beqn
\frac{dL}{dt}=[L, A_n + \bar A_n] + (B_n + \bar B_n) L.
\enqn

In other words, the system \eqref{hierarchy} was essentially a letter of acquaintance from an infinitely diverse family of novel $(1+2)$ dimensional equations. And the very first (and the simplest) member of that very family was the $n=1$ equation (with $A_1=\partial^3 + 3 \omega(x,y,t) \partial$) that has been henceforth known as the {\em Novikov-Veselov equation} (NV) \cite{Novikov84}:

\beq \label{N-V}
\begin{split}
u_t & = \partial^3 u + (\bar\partial)^3 u + 3 \left(\partial (u \omega) + \bar\partial (u \bar \omega)\right)\\
\partial u &= \bar\partial\omega.
\end{split}
\enq

Subsequent studies of NV equation has produced a lot of very interesting observation: for example, in a one-dimensional case (i.e. when both $u$ and $\omega$ are independent of $y$ variable) the \eqref{N-V} reduces to the KdV equation; whereas if we add to \eqref{N-V} one additional term, $\lambda \partial \omega$, and then take a limit $\lambda \to \pm \infty$, we will instead wound up with either one of two Kadomtsev-Petviashvili equations \cite{Zakharov91} -- another famous $(1+2)$ generalization of KdV! In fact, a huge and ever-growing body of works related to the study of NV equations has been established (see, for example, \cite{Grinevich88}, \cite{Lassas12}, \cite{Perry14} and also \cite{Croke15} for a rather extensive review of a recent literature on the subject). In particular, a lot of spotlight has been focused on the solutions of \eqref{N-V} and \eqref{Novikov84}. For example, the article \cite{1Nickel} shows how the method of superposition originally proposed in \cite{Yun} can be used to obtain a 2-solitary wave solution of the Novikov-Veselov equation. The more general N-solitons solutions were subsequently constructed in \cite{Jen}. A conspicuous absence of exponentially localized solitons for NV equation with a positive energy was explored and explained in \cite{1Novik}, whereas the impossibility of such solutions for the negative energy NV was proven in \cite{2Novik}. Many articles were dedicated to unusual and fascinating properties of the multi-dimensional solutions, including those for seemingly ordinary flat waves. In particular, in \cite{Croce1} it has been shown that plane wave soliton solutions of NV equation are not stable for transverse perturbations; the paper \cite{Croke15} demonstrates that NV equation permits such interesting solutions as multi-solitons, ring solitons, and the breathers; while the authors of \cite{Jen1} construct a Mach-type soliton of the NV equation. One of the most effective mathematical tools for studying the NV equation is the inverse scattering method. It was developed and applied in many articles, such as, for example, \cite{Niz} and \cite{Lassas12}. We must also mention an important paper \cite{Perry14}, which looked at a zero-energy Novikov-Veselov equation for the initial data of conductivity type. Taking into account that the $(1+2)$ nonlinear equations to this day remains mostly ``terra incognita'', it generates a lot of attention when someone manages to establish a relationship between the various members of a small group of currently known integrable models. As one such example we can refer to the article \cite{Zi} which has uncovered a curious relationship between the hyperbolic NV equation and the stationary Davey-Stewartson II equation -- here an adjective ``hyperbolic'' simply means that the $L$ equation \eqref{Wheeler-DeWitt-1} is hyperbolic, i.e. that both $x$ and $y$ variables are real (accordingly, since the ``original'' NV equation is associated with the system \eqref{hierarchy}, it can be called ``elliptic''). Finally, a lot of literature has been written on the subject of various generalizations of NV equations, of their properties and of their solutions \cite{L1}, \cite{L2}, \cite{L3}, and see also \cite{Konoplya}, where the analogue of NV equations is shown to arise in the nonlinear optics in a dispersion-free limit.

The observant reader will of course notice that one of the most prominent aspects of the majority of the articles we have mentioned is an almost universal adoption of an inverse scattering method as a primary tool for conducting the research and finding the exact solutions of NV. However, in this article we wish to discuss an alternative method of solving the Cauchy problem for NV (the hyperbolic version). This method, albeit simple in principle, appears to be deep enough to be applicable to a very broad class of equations, NV being just the first one -- just as it is the but a first member of the Novikov-Veselov hierarchy.

The article is organized as follows. In Sec. \ref{sec:Moutard} we introduce all the necessary ingredients of our proposed method, namely: the Lax pair for the hyperbolic (real-valued) NV equation, the Moutard transformation and the Airy functions -- and describe how to use them to produce the exact solutions to the NV equation. In the next section, Sec. \ref{Sec:Cauchy}, we up the ante by adding the initial conditions into the mix -- and show how to make sure the new solution satisfies those conditions. Finally, in Sec. \ref{sec:Higher} we discuss the generalization of the proposed method to the higher-order equations from the Novikov-Veselov hierarchy.


\section{The Moutard Transformation} \label{sec:Moutard}
Let us start by introducing the hyperbolic NV equation:
\beq \label{Novikov84}
\begin{split}
u_t & = u_{xxx}+u_{yyy}+3 \Big((a u)_x + (b u)_y\Big)\\
u_x & = a_y, \quad u_y = b_x,
\end{split}
\enq
where from now on the indices will denote the partial derivatives w.r.t. the corresponding variables. This system allows for a Lax pair of the following type:
\beq \label{LA}
\begin{split}
&\Psi_{xy} + u \Psi = 0\\
&\Psi_t = \Psi_{xxx} + \Psi_{yyy} + 3\big(a\Psi_x + b \Psi_y\big).
\end{split}
\enq

If one knows two linearly independent solutions $\Psi_1(x,y,t)$ and $\Psi_2(x,y,t)$ for \eqref{LA}, then one can utilize the famous Moutard transformation to construct a new function $\Psi[1](x,y,t)$ that will serve as a solution to the same equation \eqref{LA} albeit with a new potential $u[1](x,y,t)$. The new potential will then satisfy the relation
\beq \label{u1}
u[1] = u + 2 \partial_x \partial_y \ln \Psi_1.
\enq

Let us assume that $u=a=b=0$. Then the entire system \eqref{LA} simplifies to
\begin{align}
&\Psi_{xy} = 0 \label{LA1} \\
&\Psi_t = \Psi_{xxx} + \Psi_{yyy}. \label{LA2}
\end{align}
The equation \eqref{LA1} can be resolved by separating the variables. The resulting solution will be of a form:
\beq \label{psi1}
\Psi_1(x,y,t)=A(x,t)+B(y,t),
\enq
where $A,B$ are two arbitrary functions that are continuously differentiable by $x$ and $y$, correspondingly. Substituting \eqref{psi1} into \eqref{u1} yields a following post-Moutard form of function $u[1](x,y,t)$:
\beq \label{u}
u[1]=-2\frac{ \partial_x A \cdot \partial_y B}{(A+B)^2}.
\enq
As follows from \eqref{u}, our next goal should lie in ascertaining the exact forms of the functions $A(x,t)$ and $B(y,t)$. This task can be accomplished by looking at the equation \eqref{LA2} which we have ignored so far. We will rewrite it as a standard Cauchy problem by introducing the initial conditions for $A(t,x), B(t,y)$
\beq \label{init}
A(0,x) = \phi(x), \qquad B(0,y) = \Phi(y).
\enq
and rewriting the \eqref{LA2} as a system
\begin{equation} \label{AB}
\begin{split}
A_t & = A_{xxx} + T(t)\\
B_t & = B_{yyy} - T(t),
\end{split}
\end{equation}
where $T = T(t)$ is an arbitrary time-dependent function. The apparently symmetric nature of \eqref{AB} allows us to restrict our attention on just one of the equations therein, namely -- the first one.

We begin by introducing the Fourier transform $\tilde A(p,t)$ of the function $A(x,t)$:
\begin{equation*}
\tilde A(p,t)=\frac{1}{\sqrt{2 \pi}} \int\limits^\infty_{-\infty} A(x,t) \eb^{-i p x} dx.
\end{equation*}

This transformation is handy because of the identity
\beq \label{Afourier}
A(x,t)=\frac{1}{\sqrt{2 \pi}} \int\limits^\infty_{-\infty} \tilde A(p,t) \eb^{i p x} dp,
\enq
which, after being substituted into \eqref{AB}, yields the equation
\begin{equation} \label{ApI}
\int\limits^\infty_{-\infty} \left(\frac{\partial \tilde A}{\partial t} + i p^3 \tilde A - T \right) \eb^{i p x} dp = 0.
\end{equation}
The equation \eqref{ApI} must be satisfied for all $x$ and $p$, and therefore leads to:
\begin{equation} \label{Ap}
\frac{\partial \tilde A}{\partial t} + i p^3 \tilde A = T(t).
\end{equation}

\eqref{Ap} is a nonhomogeneous linear O.D.E. of first order. Its general solution is
\beq \label{Apsol}
\tilde A(p,t)=C(p) \eb^{-i p^3 t} +\int\limits_0^t T(\tau)  \eb^{-i p^3(t-\tau)} d\tau,
\enq
where $C(p)$ is a function, determinable from the initial conditions \eqref{init}. Using the inverse Fourier transform \eqref{Afourier} we come to the following conclusion:
\beq \label{AwithC}
A(x,t)=\frac{1}{\sqrt{2\pi}} \int\limits_{-\infty}^\infty{\left(\int\limits_0^t T(\tau) \eb^{i p^3 \tau} d\tau +C(p) \right)} \eb^{i p x-i p^3 t} dp.
\enq

According to \eqref{init},
\beqn
\phi(x)=\frac{1}{\sqrt{2 \pi}} \int\limits_{-\infty}^\infty C(p) \eb^{i p x} dp,
\enqn
so the unknown $C(p)$ is an inverse Fourier transform of the initial condition $\phi(x)$, i.e.:
\beqn
C(p)=\frac{1}{\sqrt{2 \pi}} \int\limits_{-\infty}^\infty \phi(x) \eb^{-i p x} dx.
\enqn
and we subsequently end up with the following general formula for the function $A(x,t)$:
\beq \label{A}
A(x,t)= \frac{1}{\sqrt{2 \pi}} \int\limits_0^t T(\tau) d\tau \int\limits_{-\infty}^\infty \eb^{i p x - ip^3 (t-\tau)} dp + \frac{1}{2 \pi} \int\limits_{-\infty}^\infty \phi(\xi) d\xi \int\limits_{-\infty}^\infty \eb^{i p (x - \xi) - i p^3 t} dp.
\enq

The \eqref{A} can be further simplified by pointing out the similarity between the integrals with respect to variable $p$ and the {\em Airy function} $\Ai(\xi)$. The Airy function is a particular solution of the eponymous Airy equation:
\beq \label{AiryEq}
\frac{d^2 z}{d \xi^2} = \xi z,
\enq
that has a following integral representation:
\beq \label{Airy}
\Ai(\xi) = \frac{1}{\sqrt{2\pi}} \int\limits_{-\infty}^\infty \eb^{i\left(\frac{t^3}{3}+\xi t\right)} dt.
\enq

Using this fact together with the apparent identity:
\beqn
\frac{1}{\sqrt{2\pi}} \int\limits_{-\infty}^\infty \eb^{-i p a - i p^3 b} dp = \frac{1}{\sqrt[3]{3b}} ~\Ai\left(\frac{a}{\sqrt[3]{3b}}\right),
\enqn
together with the equation \eqref{A} and the similar one written for $B(y,t)$ finally yields:
\beq
\begin{split} \label{ABsol}
A(x,t) &= \int\limits_0^t \frac{T(\tau)}{\sqrt[3]{3(\tau-t)}} ~\Ai\left(\frac{x}{\sqrt[3]{3(\tau-t)}}\right) d\tau + \frac{1}{\sqrt{2\pi}\sqrt[3]{3t}}\int\limits_{-\infty}^\infty \phi(\xi) \Ai \left(\frac{\xi-x}{\sqrt[3]{3t}}\right) d\xi \\
B(y,t) &= - \int\limits_0^t \frac{T(\tau)}{\sqrt[3]{3(\tau-t)}} ~\Ai\left(\frac{y}{\sqrt[3]{3(\tau-t)}}\right) d\tau + \frac{1}{\sqrt{2\pi}\sqrt[3]{3t}}\int\limits_{-\infty}^\infty \Phi(\eta) \Ai \left(\frac{\eta-y}{\sqrt[3]{3t}}\right) d\eta.
\end{split}
\enq

So, we end up with both the solution $\Psi_1=A+B$ of the Lax pair \eqref{LA1}, \eqref{LA2}, and, as a courtesy of Moutard transform \eqref{u1}, with a solution $u[1]$ of the NV equation \eqref{Novikov84} as well. In other words, to find a non-zero solution of the NV equation, it will suffice to start with $u \equiv 0$, impose the boundary conditions \eqref{init} on the Lax pair \eqref{LA1}, \eqref{LA2}, use \eqref{ABsol} to find its solution and conclude the calculations by finding a function $u[1]$ via the Moutard transformation \eqref{u1}. As straightforward as it is, there is one question we should ask: what would happen should we try to {\em invert} the process and instead start out with he boundary conditions for the NV equation itself?

\section{The Cauchy problem for the Novikov-Veselov equation} \label{Sec:Cauchy}

In the previous chapter we have shown that there shall exist a solution $u[1](x,y,t)$ to the NV equation, whose exact form can be derived via the Moutard transformation \eqref{u} from the solutions of the system (\ref{LA1}, \ref{LA2}), provided we are given the initial conditions \eqref{init}. But what would happen if the exact forms of the functions $\phi(x)$ and $\Phi(y)$ are {\em unknown} and we are instead given the initial conditions for the NV equation itself, and would it still be possible to find the required $u[1]$? In other words, is it possible to find an analytic solution to the Cauchy problem for the NV equation provided we only know that the solution has a general structure \eqref{u}? In short, the answer is ``yes''.

Let us start by introducing the set of initial and boundary conditions for the NV equation:
\beq \label{initNV}
\begin{split}
u[1](x,y,0) &= u_0(x,y)\\
u_0(x,0) &= A_1(x) \\
u_0(0,y) &= B_1(y) \\
A_1(0) &= B_1(0) = C,
\end{split}
\enq
where $C \in \R$ is some constant that is given to us together with the boundary conditions $A_1$ and $B_1$. Since we know that $u[1]$ satisfies the Moutard transformation, we also know that:
\beq \label{u0}
u_0(x,y) = -2 \frac{\phi'(x) \cdot \dot\Phi(y)}{(\phi(x)+\Phi(y))^2},
\enq
where $\phi$ and $\Phi$ are defined as in Sec \ref{sec:Moutard}, and $'$ and $\cdot$ denote the partial derivatives with respect to $x$ and $y$ variables correspondingly. From \eqref{u0} and \eqref{initNV} it immediately follows that
\beq \label{A1C}
\begin{split}
A_1(x) &= -2 \frac{\phi'(x) \cdot \dot \Phi(0)}{(\phi(x)+\Phi(0))^2}\\
B_1(y) &= -2 \frac{\phi'(0) \cdot \dot \Phi(y)}{(\phi(0)+\Phi(y))^2} \\
C &= -2 \frac{\phi'(0) \cdot \dot \Phi(0)}{(\phi(0)+\Phi(0))^2}.
\end{split}
\enq
The first two differential equations in \eqref{A1C} can be easily integrated; for example, the first one after the integration with respect to the variable $x$ yields
\beqn
\frac{\dot \Phi(0)}{\phi(x)+\Phi(0)} = \frac{1}{2}\int\limits_0^x A_1(\xi) d\xi + \frac{\dot \Phi(0)}{\phi(0)+\Phi(0)},
\enqn
which leads us to the following conclusion:
\beq \label{phi}
\begin{split}
\phi(x) &= \displaystyle{\frac{2 \dot\Phi_0}{\int\limits_0^x{A_1(\xi)d\xi} + \frac{2 \dot\Phi_0}{\phi_0+\Phi_0}}}-\Phi_0\\
\phi'(x) &= \displaystyle{\frac{-2 \dot\Phi_0 A_1(x)}{\left(\int\limits_0^x{A_1(\xi)d\xi} + \frac{2 \dot\Phi_0}{\phi_0+\Phi_0}\right)^2}},
\end{split}
\enq
where we have introduced the notation: $\phi(0)=\phi_0$, $\Phi(0)=\Phi_0$, $\phi'(0)=\phi'_0$ and $\dot\Phi(0)=\dot\Phi_0$. In a similar fashion, the boundary condition $\Phi(y)$ and its derivative will satisfy:
\beq \label{Phi}
\begin{split}
\Phi(y) &= \displaystyle{\frac{2 \phi'_0}{\int\limits_0^y{B_1(\zeta)d\zeta} + \frac{2 \phi'_0}{\phi_0+\Phi_0}}}-\phi_0\\
\dot\Phi(y) &= \displaystyle{\frac{-2 \phi'_0 B_1(y)}{\left(\int\limits_0^y{B_1(\zeta)d\zeta} + \frac{2 \phi'_0}{\phi_0+\Phi_0}\right)^2}}.
\end{split}
\enq
The system (\ref{phi}, \ref{Phi}) depends on four constants: $\phi_0$, $\Phi_0$, $\phi'_0$ and $\dot\Phi_0$. Three of them can be chosen arbitrarily, whereas the fourth one would have to satisfy the equation \eqref{A1C}, namely:
\beqn
-2 \frac{\phi'_0 \cdot \dot \Phi_0}{(\phi_0+\Phi_0)^2} = C.
\enqn
Curiously, this choice does not affect the Cauchy problem of the NV equation in the least, for it can be shown by direct substitution into \eqref{u0} that:
\beq \label{u0last}
u_0(x,y)=\frac{4 C A_1(x) B_1(y)}{\left(\int\limits_0^x \int\limits_0^y A_1(\xi) B_1(\zeta) d\zeta d\xi + 2 C\right)^2},
\enq
i.e. the {\em initial} condition $u_0(x,y)$ depends only on the known {\em initial boundary} conditions $A_1(x)$, $B_1(y)$ and $C$.

We are now ready to answer the question posed in the beginning of this section: provided we know the initial conditions \eqref{u0last}, how do we solve the corresponding Cauchy problem of the hyperbolic real-valued Novikov-Veselov equation? The answer lies in {\em repeating} the Moutard transformation process we described in Sec. \ref{sec:Moutard}! Indeed, since the unknown functions $\phi(x)$ and $\Phi(y)$ satisfy the relations \eqref{phi} and \eqref{Phi}, all we really have to do is substitute them into the system \eqref{ABsol}, derive $A(x,t)$ and $B(y,t)$, and substitute them in equation \eqref{u} to find out the sought after $u[1](x,y,t)$, which will conclude the problem.

Lets summarize everything we have said so far. In order to find an exact solution $u(x,y,t)$ to the hyperbolic real-valued Novikov-Veselov equation
\beq \label{NV1}
u_t = u_{xxx}+u_{yyy},
\enq
with the given initial boundary conditions
\beqn
u(x,0,0) = A_1(x), \qquad u(0,y,0) = B_1(y), \qquad u(0,0,0) = C,
\enqn
that correspond to the initial condition
\beqn
u_0(x,y)=\frac{4 C A_1(x) B_1(y)}{\left(\int\limits_0^x \int\limits_0^y A_1(\xi) B_1(\zeta) d\zeta d\xi + 2 C\right)^2},
\enqn
one shall:
\begin{enumerate}

\item Choose a differentiable function $T(t)$ and four numbers $\alpha$, $\beta$, $\gamma$ and $\delta$ that satisfy the condition,
\beqn
-2 \frac{\gamma \cdot \delta}{(\alpha+\beta)^2} = C.
\enqn

\item Find two support function $\phi(x)$ and $\Phi(y)$ via the formulas
\beqn
\begin{split}
\phi(x) &= \displaystyle{\frac{2 \delta}{\int\limits_0^x{A_1(\xi)d\xi} + \frac{2 \delta}{\alpha+\beta}}}-\beta\\
\Phi(y) &= \displaystyle{\frac{2 \gamma}{\int\limits_0^y{B_1(\zeta)d\zeta} + \frac{2 \gamma}{\alpha+\beta}}}-\alpha.
\end{split}
\enqn

\item Substitute $\phi(x)$ and $\Phi(y)$ into the equations
\beq \label{AB1}
\begin{split}
A(x,t) &= \int\limits_0^t \frac{T(\tau)}{\sqrt[3]{3(\tau-t)}} ~\Ai\left(\frac{x}{\sqrt[3]{3(\tau-t)}}\right) d\tau + \frac{1}{\sqrt{2\pi}\sqrt[3]{3t}}\int\limits_{-\infty}^\infty \phi(\xi) \Ai \left(\frac{\xi-x}{\sqrt[3]{3t}}\right) d\xi \\
B(y,t) &= - \int\limits_0^t \frac{T(\tau)}{\sqrt[3]{3(\tau-t)}} ~\Ai\left(\frac{y}{\sqrt[3]{3(\tau-t)}}\right) d\tau + \frac{1}{\sqrt{2\pi}\sqrt[3]{3t}}\int\limits_{-\infty}^\infty \Phi(\eta) \Ai \left(\frac{\eta-y}{\sqrt[3]{3t}}\right) d\eta.
\end{split}
\enq

\item Substitute the new functions $A(x,t)$ and $B(y,t)$ into the equation
\beq \label{uu}
u = -2\frac{ \partial_x A \cdot \partial_y B}{(A+B)^2}.
\enq
\end{enumerate}

The resulting function $u(x,y,t)$ will be a proper solution of the Cauchy problem since by construction it will satisfy both the NV equation \eqref{NV1}, and the initial conditions $u(x,y,0)=u_0(x,y)$. We would like to emphasize here that this procedure does not involve anything more complicated than partial differentiation and integration and can therefore be used for both the analytic study of the properties of the solutions of NV equation and the corresponding numerical calculations.

Before we conclude this section, we would like to offer two interesting examples of Cauchy problems that might be used for the algorithm we have described above and that serve as the proof that even the seemingly simple case of \eqref{LA1} with $u=0$ can lead to some rather interesting problems.

\begin{figure}
\begin{center}
\includegraphics[width=0.9\columnwidth]{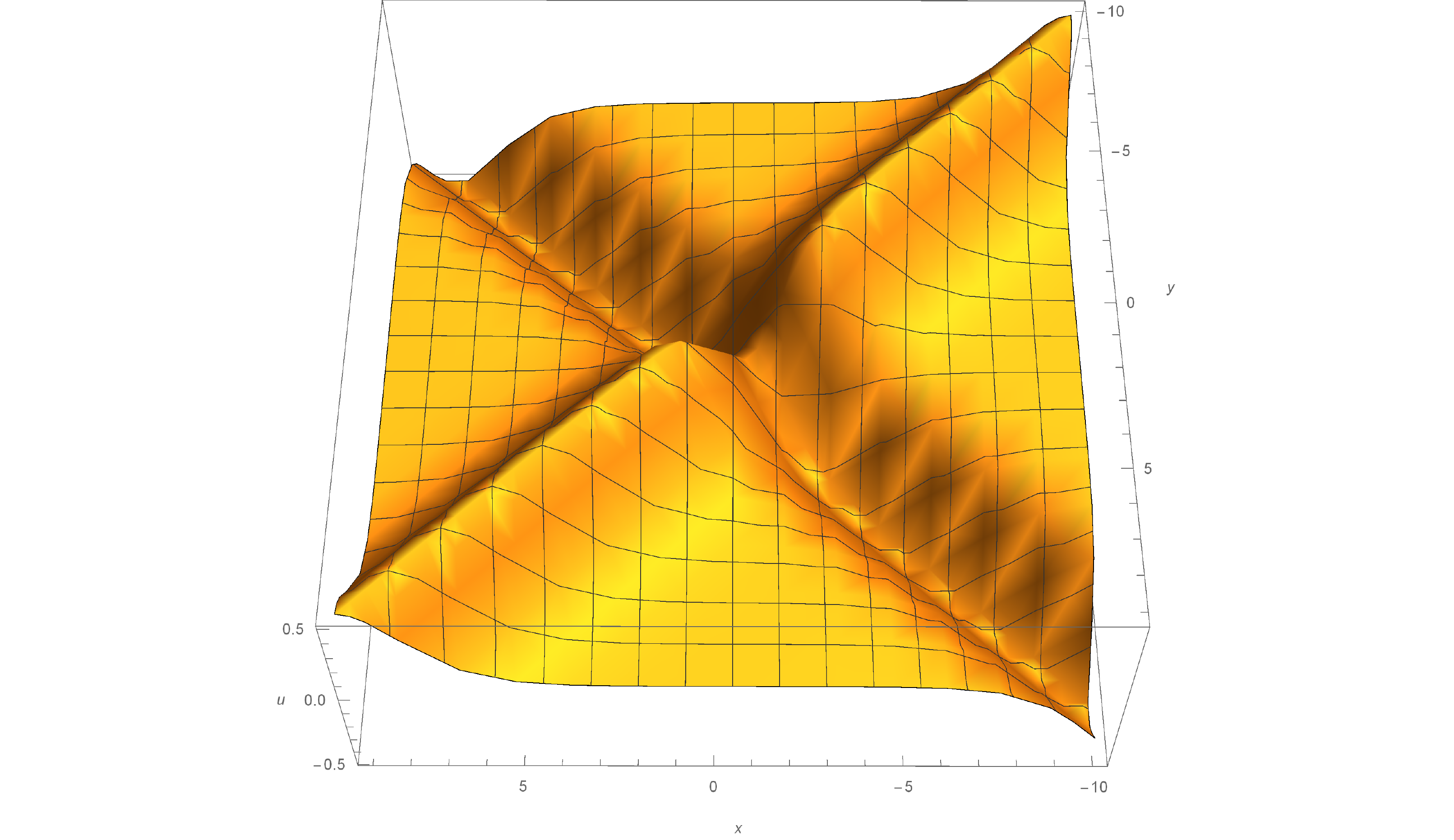}
\caption{\label{fig1} The dromion-type Cauchy problem \eqref{dromion} for $u_0(x,y)$ with $x_0 = y_0 = -1$. There are two equipotential lines at $y - y_0=\pm (x - x_0)$.}
\end{center}
\end{figure}

The first is based on the solution of \eqref{psi1} of the form: $\Psi_1(x,y,0) = \cosh (x-x_0) + \cosh (y-y_0)$, where $x_0, y_0 \neq0$. According to \eqref{u} it corresponds to the following {\em dromion solution}:
\beq \label{dromion}
u_0(x,y)=\frac{2 \sinh (x-x_0) \cdot \sinh (y-y_0)}{(\cosh (x-x_0) + \cosh (y-y_0))^2},
\enq
depicted on Fig. \ref{fig1}. It is important to note that $x_0$ and $y_0$ must be non-zero constants, otherwise the functions $\phi(x)=u_0(x,0)$ and $\Phi(y)=u_0(0, y)$ in \eqref{AB1} will be identically zero. Any other choice for $x_0, y_0 \in \R$, however, would be fine and will result in nontrivial solutions of the NV equation. Also, a little note is in order. The solution we have just produced is the exponentially exponentially localized soliton localized on the 2D plane. The solutions of this have been previously constructed for the Davey-Stewartson-I (DS) equation in \cite{Boiti}, \cite{Fokas} and \cite{Tema}, while the term ``dromion'' itself stems from the 1989 paper by Fokas ans Santini. \cite{Fokas1}. The DS equations describe two interacting fields, and in the case of the dromion on of them describes a certain exponentially localized (on a plane) structure, while the other has orthogonal equipotential lines. It is thanks to this very property that we are at liberty to call the solution \eqref{dromion} a ``dromion''.

\begin{figure}
\begin{center}
\includegraphics[width=0.9\columnwidth]{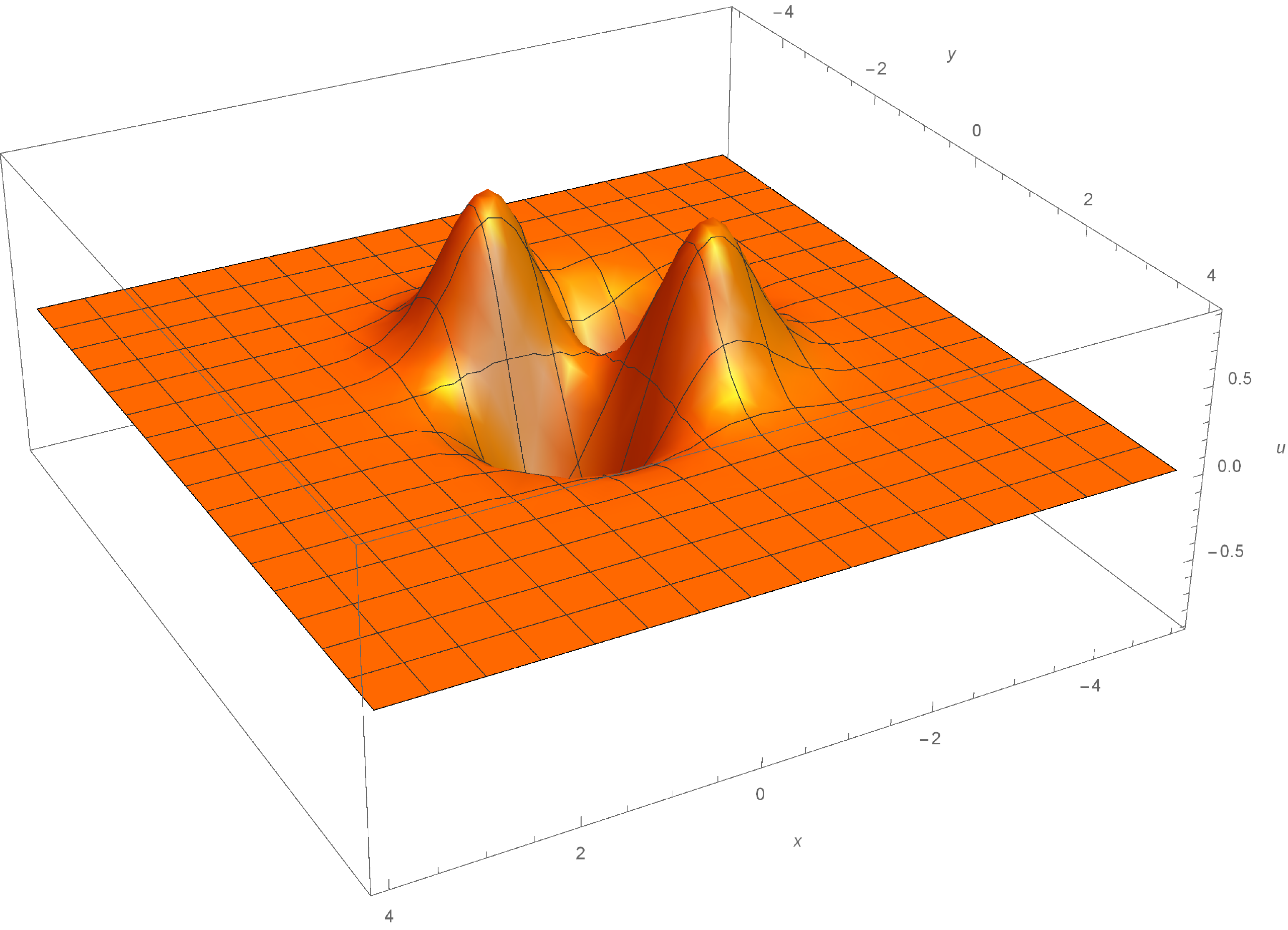}
\caption{\label{fig2} A sample of localized Cauchy problem \eqref{localized} with $x_0 = y_0 = -1$.}
\end{center}
\end{figure}

The second solution is based on the function: $\Psi_1(x, y, 0) = \eb^{(x - x_0)^2} + \eb^{-(y - y_0)^2}$, where $x_0, y_0 \neq0$. The solution $u_0(x,y)$ then has the form:
\beq \label{localized}
u_0(x,y)=\frac{8 \eb^{r^2} (x - x_0) (y - y_0)}{(1+\eb^{r^2})^2}, \qquad r^2 = (x - x_0)^2 + (y - y_0)^2.
\enq
depicted on Fig. \ref{fig2}, and it is easy to see that on the plane $0xy$ it describes an exponentially localized structure, centered around the point $(x_0, y_0)$.

\section{Generalization of the method: the higher-order equations} \label{sec:Higher}

Let us now say a few words about the more general problem. First of all, let us look at the following operator-type Lax pair:
\beq \label{LAH}
\begin{split}
&\partial_x \partial_y \Psi + u \Psi = 0\\
&\partial_t \Psi = \partial_x^n \Psi + \partial_y^n \Psi,
\end{split}
\enq
where $n \in \N^{+}$ in some non-zero natural number. This system will correspond to a family of Lax equations, with the special case $n=3$ corresponding to the hyperbolic NV equation. It will still allow for the Moutard transformation, and therefore the crux of our discussion would still be applicable for arbitrary $n$. However, one thing that {\em must} change is the exact form of the equations for $A(x,t)$ and $B(y,t)$. This is due to a very simple reason: the Fourier transform $\tilde A(p,t)$ for the function $A(x,t)$ had to satisfy the equation \eqref{Ap}, where one of the terms had a specific factor $i p^3$. It is this  very factor that had allowed us to invoke the Airy function and with its aid derive the exact formula for $A(x,t)$. Unfortunately, this very factor is endemic to the Veselov-Novikov equation, and in the more general case of \eqref{LAH} it must be replaced with the factor $-(i p)^n$ -- destroying any relationship with the Airy function proper. As a result, the entire formula \eqref{AB1} becomes no longer applicable for the general case and thus should be properly replaced. In order to find out the suitable replacement, we shall separately consider two alternative cases: when $n$ is odd and when $n$ is even.
\newline

{\bf Case 1: Odd $n$.} Let $n=2m+1$, where $m \ge 0$. Following our previous discussion, let us consider the special case $u \equiv 0$. Then the system \eqref{LAH} turns into
\beqn
\begin{split}
&\partial_x \partial_y \Psi = 0\\
&\partial_t \Psi = \partial_x^n \Psi + \partial_y^n \Psi.
\end{split}
\enqn
Since the first equation requires that $\Psi=A(x,t)+B(y,t)$, the system subsequently splits into the following equations:
\beqn
\partial_t A_{2m+1} =\partial_x^{2m+1} A_{2m+1}, \qquad \partial_t B_{2m+1} = \partial_y^{2m+1} B_{2m+1},
\enqn
where for simplicity we have omitted the arbitrary function $T(t)$. Using the Fourier transformation
\beqn
\tilde A_{2m+1}(p,t)=\frac{1}{\sqrt{2 \pi}} \int\limits^\infty_{-\infty} A_{2m+1}(x,t) \eb^{-i p x} dx,
\enqn
we end up with the differential equation
\beqn
\partial_t \tilde A_{2m+1} = (i p)^{2m+1} \tilde A_{2m+1} = i (-1)^m p^{2m+1} \tilde A_{2m+1}.
\enqn
Solving it and returning back to $A(x,t)$ as described in Sec.\ref{sec:Moutard} yields
\beq \label{Aodd}
A_{2m+1}(x,t) = \frac{1}{\sqrt{2\pi}} \int\limits_{-\infty}^{\infty} d\xi ~\phi(\xi) ~\frac{1}{\sqrt{2\pi}} \int\limits_{-\infty}^{\infty} dp ~\eb^{i\left(p(x-\xi)+(-1)^m p^{2m+1} t\right)}.
\enq

As we know, in the special case $m=1$ (i.e. $n=3$) the inner integral in \eqref{Aodd} can be rewritten in terms of the Airy function
\beqn
\Ai(\xi) = \frac{1}{\sqrt{2\pi}} \int\limits_{-\infty}^\infty \eb^{i\left(\frac{t^3}{3}+\xi t\right)} dt = \sqrt{\frac{2}{\pi}} \int\limits_0^\infty \cos{\left(\xi t +\frac{t^3}{3}\right)} dt,
\enqn
which serves as a solution to the Airy equation
\beqn
\frac{d^2 z}{d \xi^2} = \xi z,
\enqn
and is easily derived using either Fourier or Laplace transformation; in case of the Laplace transformation the contour of integration must be chosen lying inside of a sector where the polar angle $\theta$ satisfies the condition $\cos(3\theta) >0$.

Similarly, it is easy to show that one of a solutions to a more general equation
\beqn
\frac{d^{2m} z}{d \xi^{2m}} = \xi z,
\enqn
will be a {\em higher-order generalization} of the Airy function:
\beq \label{Airyodd}
\Ai_{2m+1}(\xi) = \frac{1}{\sqrt{2\pi}} \int\limits_{-\infty}^\infty \eb^{i\left(\xi t - \frac{(-1)^m}{2m+1} ~t^{2m+1}\right)} dt = \sqrt{\frac{2}{\pi}} \int\limits_0^\infty \cos{\left(\xi t - \frac{(-1)^m}{2m+1} ~t^{2m+1}\right)} dt,
\enq
which means that the required functions $A_{2m+1}$ and $B_{2m+1}$ can be derived from the initial conditions $\phi(x)$ and $\Phi(y)$ by the following formulas:
\beq \label{ABodd}
\begin{split}
A_{2m+1}(x,t) &= \frac{1}{\sqrt[2m+1]{(2m+1)t}} \int\limits_{-\infty}^\infty d\xi ~\phi(\xi) ~\Ai_{2m+1}\left(\frac{\xi-x}{\sqrt[2m+1]{(2m+1)t}}\right),\\
B_{2m+1}(x,t) &= \frac{1}{\sqrt[2m+1]{(2m+1)t}} \int\limits_{-\infty}^\infty d\zeta ~\Phi(\zeta) ~\Ai_{2m+1}\left(\frac{\zeta-y}{\sqrt[2m+1]{(2m+1)t}}\right).
\end{split}
\enq
\newline
{\em SIDE NOTE.} We would like to remind the reader that in literature the term {\em generalized Airy function} is commonly assigned to the solutions of the second order O.D.E. $w''(x)=x^n w(x)$; hence the addition of the term {\em
-order} in our case is necessary to avoid a possible confusion.
\newline

{\bf Case 2: Even $n$.} Let $n=2m$, where $m \ge 0$. This time let us utilize not a Fourier but a Laplace transform:
\beqn
A_{2m}(x,t)=\int\limits^\infty_{-\infty} \tilde A_{2m}(p,t) \eb^{p x} dx,
\enqn
where we have introduced the equation for $\tilde A(p,t)$ is
\beqn
\frac{\partial \tilde A_{2m}}{\partial t} = p^{2m} \tilde A_{2m},
\enqn
so the required function $A(x,t)$ will satisfy the equation
\beq \label{Aodd}
A_{2m}(x,t) = \int\limits_{-\infty}^{\infty} d\xi ~\phi(\xi) ~\int\limits_{-\infty}^{\infty} dp ~\eb^{p(x-\xi)+p^{2m} t}.
\enq

It is not difficult to show that the Laplace transformation method applied to the ordinary differential equation
\beqn
\frac{d^{2m-1} z}{d \xi^{2m-1}} = \xi z,
\enqn
will yield a following solution
\beq \label{Airyeven}
\Ai_{2m}(\xi) = \int\limits_{-\infty}^\infty \exp \left(\xi t - \frac{t^{2m}}{2m}\right) dt,
\enq
and so the even case produces the formulas that are quite similar to the old ones, namely:
\beq \label{ABeven}
\begin{split}
A_{2m}(x,t) &= \frac{1}{\sqrt[2m]{-2 m t}} \int\limits_{-\infty}^\infty d\xi ~\phi(\xi) ~ \Ai_{2m}\left(\frac{x-\xi}{\sqrt[2m]{-2mt}}\right),\\
B_{2m}(x,t) &= \frac{1}{\sqrt[2m]{-2mt}} \int\limits_{-\infty}^\infty d\zeta ~\Phi(\zeta) ~\Ai_{2m}\left(\frac{y-\zeta}{\sqrt[2m]{-2mt}}\right).
\end{split}
\enq
Note the appearance of a negative sign under the root in \eqref{ABeven}, which serves as a indication of an ill-posedness of our problem for $t>0$.

{\em SIDE NOTE} The choice of the lower limit in the Laplace transform being equal to $-\infty$ is neither random nor capricious -- it is necessary so that during the integration by parts (which is required to get the solution \eqref{Airyeven}) all the boundary terms will vanish.
\newline

In order to conclude this section, we have to produce a sample of a differential equation that allows for a Moutard transformation and whose Lax pair reduces to \eqref{LAH} when $u=0$. It is possible to demonstrate that no such equation exists for $n=4$, which gives a very strong indication that the method is intricately  tied up to the Novikov-Veselov hierarchy. And indeed, if we consider the following Lax pair:
\beq\label{Lax_5_order}
\begin{split}
\Psi_{xy} &= - u \Psi \\
\Psi_t &= \Psi_{5x} + \Psi_{5y} + 5 \left(\omega_{yy} \Psi_{yy} \right)_y + 5 \left(\omega_{xx} \Psi_{xx} \right)_x + a \Psi_y + b \Psi_x,
\end{split}
\enq
with the functions $a$, $b$ and $\omega$ satisfying the conditions:
\beq\label{a_b_omega}
\begin{split}
\omega_{xy} &= u \\
a_x &= 5 u_{yyy} + 5 \left(u \omega_{yy}\right)_y + 10 u_y \omega_{yy} \\
b_y &= 5 u_{xxx} + 5 \left(u \omega_{xx}\right)_x + 10 u_x \omega_{xx},
\end{split}
\enq
then we end up with the second equation from the Novikov-Veselov hierarchy, which has the following form:
\beq \label{NS_2}
u_t = u_{5x} + u _{5y} + 5 \Big((\omega_{yy} u_y)_{yy} + (\omega_{xx} u_x)_{xx} \Big) + (a u)_y + (b u)_x,
\enq
which exactly satisfies both of our underlying assumptions whenever we set $\omega=0$, or, more generally, choose $\omega=c_1 x + c_2 y + c_3$, where $c_1, c_2, c_3 \in \R$. Subsequent application of the proposed method to this equation (plus the initial conditions) for the case $u=0$ we leave to the reader.

\section{Conclusion}\label{sec:conclusion}
At last, let us reflect on the results we have gained. The main aim of the article was to demonstrate that the Moutard symmetry is not only well-suited to construct explicit partial solutions of nonlinear partial differential equations such as Novikov-Veselov equation, but it is also versatile enough to solve the Cauchy problems like \eqref{initNV}. This is a very important and useful property of the Moutard symmetry, because the common way to solve the Cauchy problem invokes the inverse scattering method whose applications for the $(1+2)$-dimensional equations remains far from being completely understood. Another interesting application of the Moutard symmetry may be connected with the construction of so-called dressing chains of discrete symmetries. Such chains connect different integrable hierarchies and in the long run suggest that potentially all the nonlinear integrable equations might be different manifestations of one unique integrable differential equation, but simply written in different calibrations (see, for example, \cite{Yurov03}).

Another interesting open question is the evolution of initially exponentially localized structures. There is a set of theorems that explicitly forbids the existence of exponentially localized solitons of NV equation (see the discussion in Sec. \ref{sec:Intro}), so we shall expect that the evolution of such structures as \eqref{dromion} and \eqref{localized} will destroy the localization and transform the solution into some other type of structure. Our approach opens the window of opportunity to study these complex processes analytically.

\section*{References}




\end{document}